\documentclass[aps,prl,preprint,superscriptaddress]{revtex4-1}

\usepackage{graphicx}
\usepackage{bm}
\usepackage{amssymb}

\begin{document}


\title{Mechanism of small-polaron formation in the biferroic YCrO$_3$ doped with calcium}

\author{A. Dur\'an}
\email[Corresponding author: ]{dural@cnyn.unam.mx}
\affiliation{Universidad Nacional Aut\'onoma de M\'exico, Centro de Nanociencias y Nanotecnolog\'{\i}a, Apartado Postal 41, C. P. 22800, Ensenada, B. C.}
\author{E. Verd\'{\i}n}
\affiliation{Universidad de Sonora, Departamento de F\'{\i}sica, Apartado Postal 1626, Hermosillo, Sonora. C.P. 8300, M\'exico. }
\author{R. Escamilla}
\author{F. Morales}
\author{R. Escudero}
\affiliation{Universidad Nacional Aut\'onoma de M\'exico, Instituto de Investigaciones en Materiales, Apartado Postal 70-360, M\'exico D. F. 04510, M\'exico.}


\begin{abstract}
The effects of Ca substitutions on the structure, magnetism and electrical properties of YCrO$_3$ ceramics are investigated by X-ray diffraction, magnetic susceptibility and electrical conductivity measurements. The cell volume decrease occurs through the change from Cr(III) to Cr(IV) as a result of the charge compensation of the Ca doping. No changes are observed in the antiferromagnetic transition temperature while strong changes are observed in the transport measurements due to Ca content. The increase of the electrical conductivity as well as the decrease of the activation energy are caused by the formation of the small-polarons localized in the O-Cr-O lattice distortion. The origin of small-polarons in the undoped sample is different in nature from the calcium doped. ``Local non-centrosymetry" is the source of the small-polaron formation in undoped sample, while the change from Cr(III) to Cr(IV) through the charge compensation of Ca(II) in the Y(III) site is the source of small-polarons formations. The decrease of the average bond length Cr-O as well as effective moments in the paramagnetic state and the increase of the electrical conductivity are clear evidence that the Ca doping induces localized polarons, which in turn, these quasiparticles move from site to site by a thermally activated process in the doped YCrO$_3$ compound. Here, we also discuss a possible mechanism of small-polaron injections in YCrO$_3$ matrix.
\end{abstract}


\maketitle

\section{Introduction}

In the past decades many investigations were performed in orthocromites with RCrO$_3$ formula and R=Y, and rare earth (\cite{1} and reference therein, \cite{2,3,4}). The investigations were focused on understanding the magnetic and electronic properties at low temperatures. On the other hand the chemical, structural and electrical stability at high temperatures were considered promising in applications such as refractory electrodes, thermistors and thermoelectric materials \cite{5,6,7}. In recent years there has been much interest in multiferroic materials. The coupling between ferromagnetic and ferroelectric order parameters in the same phase is very attractive from both basic science and applications; {\sl i. e.} spintronic and storing data devices \cite{8,9}.

The origin of the coexistence of the ferroelectricity and feromagnetism is complex and many intrinsic features are still under debate. Recently, YCrO$_3$ has been reported as biferroic material presenting a magnetic ordering at 140 K, whereas relaxor ferroelectric behavior has been found at about 450 K. Additional characteristic of this system is that the space group is centrosymetric ({\sl Pbnm}) and, thus non-compatible with ferroelectricity. However, the occurrence of the ferroelectricity has been attributed to local non-centrosymetric nano-regions \cite{10,11}. Nevertheless, the origin of the relaxor ferroelectric transition can also be due to structural instabilities since it should explain the ferroelectricity in the RCrO$_3$ with heavy rare earth ions (R=Er-Lu) and the absence in the light ones (R=La-Tb) of the orthochromites family \cite{12}.

Ferroelectricity, and relaxor ferroelectric behavior in perovskite have showed a strong dependence with the A or B substitution sites. Thus, donor and acceptor iso- and aliovalent doping cations should play an important role in both magnetic and ferroelectric properties of these compound. In most cases, the intrinsic semiconducting behavior and the injected carrier by doping cations are detrimental for the optimal ferroic properties mainly ferroelectric hysteresis loop and dielectric loss \cite{13,14}. Among other important effects that play a significant role in the dielectric properties and electrical conductivity are the thermal history, microestructural control (random crystal orientation), grain boundary, densification, porosity, micro-cracks, etc. These factors contribute strongly to the electrical conductivity and consequently to the ferroelectric response. Contrary to this, the conductive process and the dielectric loss are strongly suppressed in single crystal or thin films samples. For instance, well defined ferroelectric hysteresis loops have been observed in single crystal and thin films of BiFeO$_3$ samples, whereas high dielectric loss and circular hysteresis loops are frequently observed in bulk ceramic samples \cite{14}. Moreover, other effects cannot be discarded; for instance, the coexistence of ferromagnetic and ferroelectric ordering must be carefully balanced by the empty or partially filled d-orbital in the octahedral B-site; and also  the interactions between charges, spin and lattice degrees of freedom. Those aspects are some of the fundamental issues in current multiferroics physics. Lastly, the new concept of ``local noncentrosymmetry" as off-centering distortion has been proposed as the origin of the small polarization in YCrO$_3$. Furthermore, this new finding supports the origin of the small- polarons as charge carriers in the YCrO$_3$ sample. This fact plus local induced lattice deformation by Ca substitutions support that the localized charge accompanying with the local polarization field plus magnetic instabilities can be moved through the crystal under electric field propagation as small-polarons.

In this work, we are addressing to the electric and magnetic behaviors observed in polycrystalline YCrO$_3$ samples doped with Ca. We found that such substitutions alter the total charge of the YCrO$_3$ matrix which is detrimental to the dielectric properties. Such substitutions are compensated for the valence change from Cr(III) to Cr(IV) which produces local distortions and in turns charge transfer in form of small-polarons. Finally, for the first time, we discuss the origin of the small-polarons in the undoped sample and the mechanism by which these quasi-particles are injected through the crystal when calcium is introduced into the YCrO$_3$ matrix.

\section{Experimental Procedure}

Polycrystalline samples were prepared by combustion method. Stoichiometric amounts of powders of Y(NO$_3$)$_3\cdot$6H$_2$O, Cr(NO$_3$)$_3\cdot$9H$_2$O, and Ca(NO$_3$)$_2$ were used as the starting materials. The synthesized compounds were Y$_{1-x}$Ca$_x$CrO$_3$ with x= 0, 0.025, 0.050, 0.075, 0.10 and 0.15. Details of combustion synthesis of YCrO$_3$ compounds are described elsewhere \cite{15}. For the structural analysis, it was used a Siemens D-5000 diffractometer with Cu K$\alpha$ radiation and Ni filter. Measurements were performed in steps of 0.02$^{\circ}$ for 14 seconds in the 2$\theta$ range of 5$^\circ$ to 120$^\circ$ at room temperature. Crystallographic phases were identified by comparison with the X-ray patterns of the JCPDS database. Structural parameters were refined by the Rietveld method using the QUANTO program with multi-phase capability \cite{16}. Magnetization was taken with a SQUID based magnetometer MPMS-5T (Quantum Design). For the dielectric measurements silver electrodes were painted on discs in order to make parallel-plate ceramic capacitors. Capacitance measurements were carried out from room temperature up to 800 K at different frequencies, in the range of 10 kHz - 1 MHz using an LCR bridge (HP-4284A).

\begin{figure}
\includegraphics[scale=0.5]{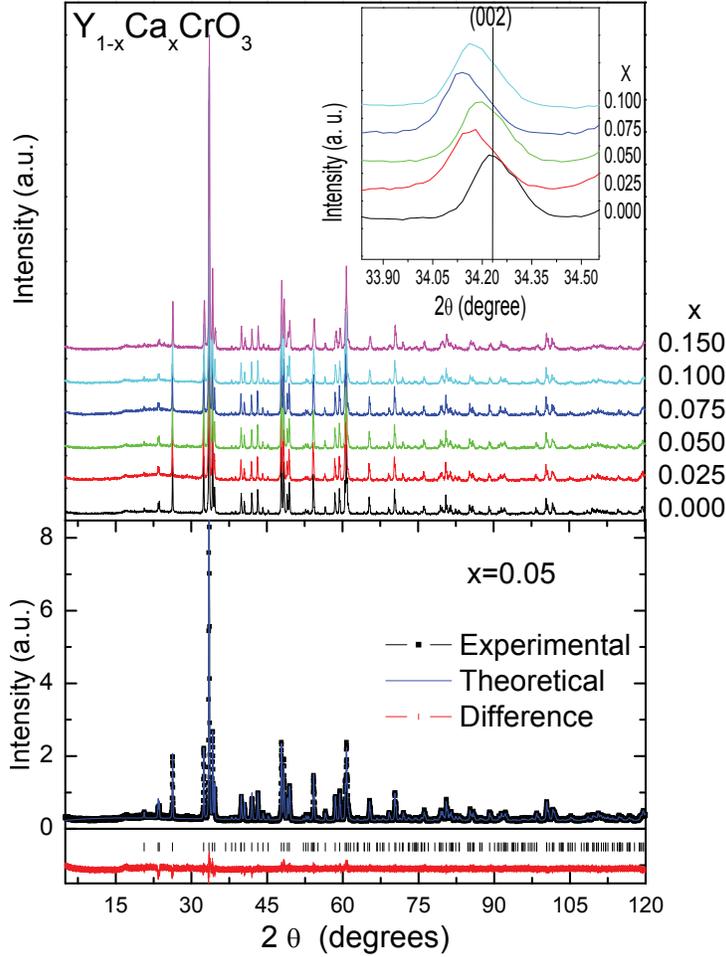}%
 \caption{\label{f1}Upper panel: X-ray diffraction patterns of the Y$_{1-x}$Ca$_x$CrO$_3$ system. The inset shows the shift of the plane (0 0 2) as a function of Ca content. Bottom panel: The Fitting results of the Rietveld analysis of the powder X-ray diffraction pattern for the x=0.0 sample. Experimental ($\blacksquare$), calculated (---) patterns. The bottom line is the difference between the observed and calculated patterns.}
\end{figure}

\section{Results.}

\subsection{Structure analysis}

Figure 1 shows the X-ray diffraction patterns for the Y$_{1-x}$Ca$_x$CrO$_3$ ($0 \leq x \leq 0.15$) samples. The analysis of these data indicates that the crystal structure corresponds to YCrO$_3$ structure (ICDD n$^\circ$ 34-0365) without the presence of a second phase. It is observed a shift to low angles of the plane (0 0 2) as gradual substitution of Ca by Y in the crystal lattice (see inset of Fig. 1). The X-ray diffraction patterns of the samples were Rietveld-fitted using the space group {\sl Pbnm} (n$^\circ$ 62) taking into account that Ca occupies Y sites. The fitted X-ray diffraction pattern for the 5\% of Ca doped sample is displayed at the bottom of Fig. 1. Structural details for all samples obtained from Rietveld refinements are shown in Table I. The profile fit shows in all samples values of  $\chi^2 < 2$, confirming the well convergence to the {\sl Pbnm} space group. The lattice parameters for the pristine sample are in agreement with other published results \cite{11,15,17}. The subtle changes in the lattice parameters and the unit cell volume as Ca replaces Y sites can be more clearly seen in Fig. 2. The structural analysis reveals the average $\langle$Cr-O$\rangle$ distance, the octahedral distortion ($\Delta$), and the tilt away from the c-axis, This is referred as $\langle\phi\rangle$, (calculated and tabulated in Table II). These internal structural parameters are essential for understanding the electric transport and magnetic exchange interaction, as will be seen below. Moreover, the results indicate that charge compensation in the environment of Cr does not affect the tilting away of c-axis.

\begin{table}
 \caption{\label{t1}Structural parameters and atomic positions for (Y$_{1-x}$Ca$_x$)CrO$_3$ system at room temperature.}
\begin{ruledtabular}
\begin{tabular}{ccccccccc}
   & x= & &0.00 & 0.025 & 0.050 & 0.075 & 0.100 & 0.150 \\
   \hline
   & a (\AA) & &5.2437(3) & 5.2467(3) & 5.2445(1) & 5.2465(1) & 5.2477(2) & 5.2499(2) \\
   &b (\AA) & &5.5235(3) &  5.5221(2) &  5.5196(1)  & 5.5124(1) &  5.5090(2)  & 5.5084(3) \\
    &c (\AA) & &  7.5360(2) &  7.5366(3) &  7.5337(2) &  7.5301(2) &  7.5287(3) &  7.5287(3) \\
   & V(\AA$^3$)& &  218.270& 218.356& 218.082& 217.777& 217.652& 217.719 \\
Y  & &x &  -0.0190(3)&  -0.0174(4) & -0.0173(3) & -0.0168(4) & -0.0165(2) & -0.0166(2)\\
   && y &  0.0680(1) &  0.0660(2) &  0.0660(1) &  0.0652(2) &  0.0659(2) &  0.0646(2) \\
    &&B(\AA$^2$) &  0.20(3)& 0.37(4)& 0.36(3)& 0.47(3)& 0.31(4)& 0.39(5)\\
Cr & & B(\AA$^2$) &  0.26(3)& 0.22(3)& 0.29(3)& 0.26(5)& 0.36(3)& 0.46(3)\\
O(1)  &&  x &  0.111(2) &   0.101(1) &   0.098(2) & 0.096(2)&    0.096(2)&    0.096(2)\\
&&    y &  0.461(1) & 0.466(2)  & 0.466(1)&    0.469(2)  &  0.466(1)&    0.469(1)\\
  &&  B(\AA$^2$) &  1.24(3)& 1.46(3)& 1.70(2) &1.64(1) &1.46(3)& 1.20(3)\\
O(2) &&   x &  -0.307(1) &  -0.308(1) &  -0.307(1) &  -0.307(2) &  -0.306(1)  & -0.307(2)\\
 &&   y &   0.307(1)&    0.301(1)&    0.301(1)&    0.302(2) &   0.301(1) &   0.301(2)\\
  &&  z  &  0.058(1) &   0.054(1) &   0.053(1) &   0.052(1) &   0.053(1)  &  0.052(1)\\
&&    B(\AA$^2$) &  1.30(2)& 1.35(2)& 1.24(5)& 1.39(5) &1.51(3)& 1.29(5)\\
\hline
&& R$_p$(\%) & 5.8 & 5.6 & 5.2 & 5.7 & 5.1 & 5.9 \\
&& R$_{wp}$(\%) & 7.3 & 6.9 & 6.4 & 7.1 & 6.4 & 7.7 \\
&& R$_{exp}$(\%) & 5.2 & 5.8 & 5.3 & 5.0 & 5.0 & 5.1 \\
&& $\chi^2$(\%) & 1.4 & 1.2 & 1.2 & 1.4 & 1.3 & 1.5 \\
\end{tabular}
\end{ruledtabular}
Note. Space group: $Pbnm$.  Atomic positions: Y: 4c  (x, y, 0.25); Cr: 4b (0, 0.5, 0); O(1): \\
4c (x, y, 0.25) and O(2):  8d (x, y, z).
\end{table}

\begin{figure}
\includegraphics[scale=0.5]{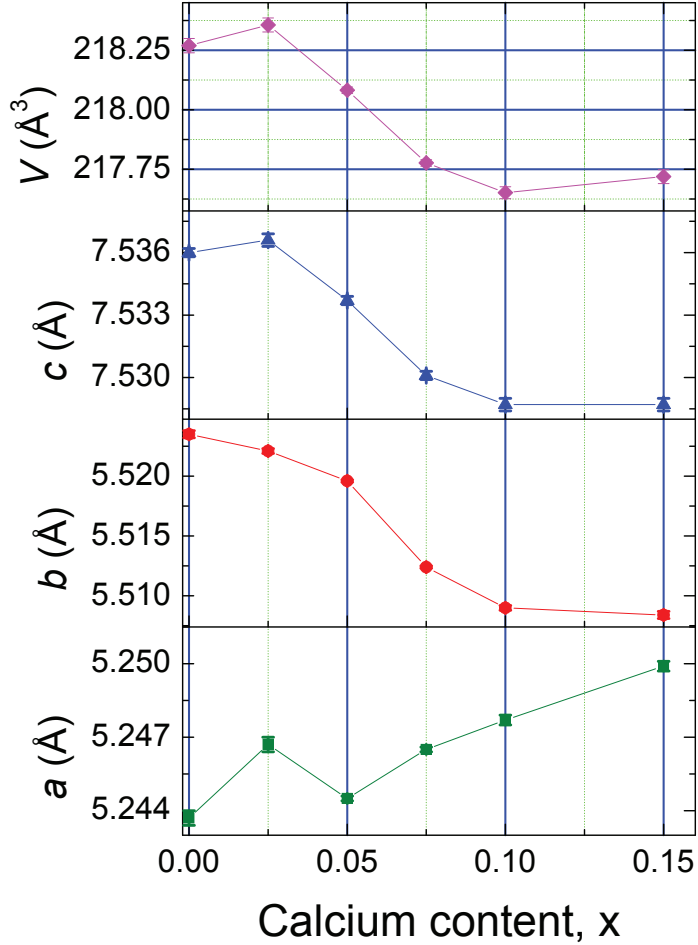}
\caption{\label{f2}Lattice parameters of the orthorhombic structure and unit cell volume as a function of Ca content in YCrO$_3$.}
\end{figure}

\begin{figure}
\includegraphics[scale=0.5]{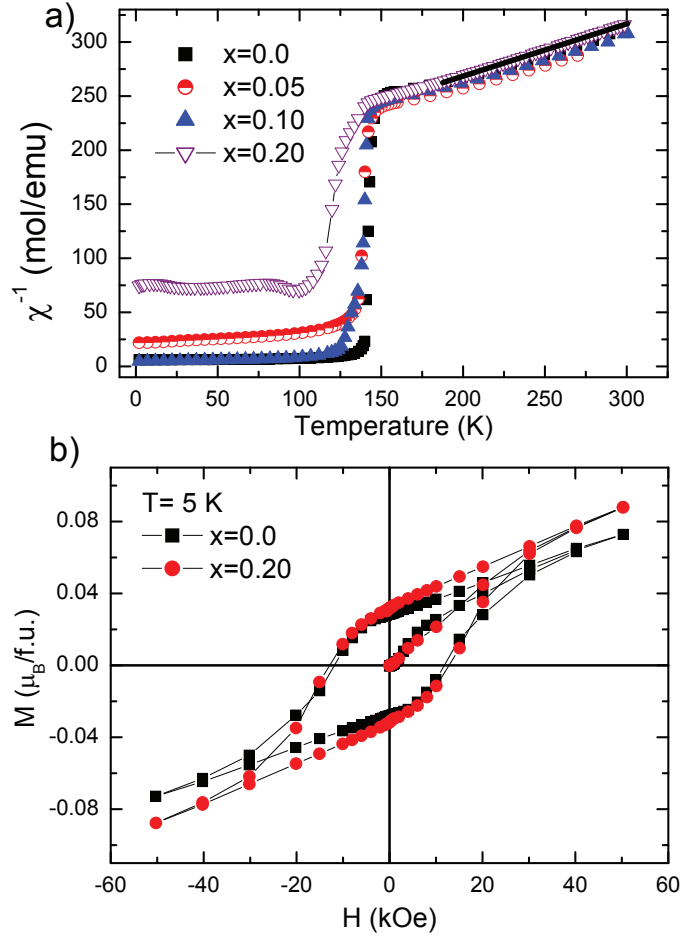}
\caption{\label{f3}Temperature dependence of the inverse susceptibility for Y$_{1-x}$Ca$_x$CrO$_3$ with $0\leq x \leq 0.15.$}
\end{figure}

\begin{table}
\caption{\label{t2}Geometrical parameters characterizing the crystal structure of (Y$_{1-x}$Ca$_x$)CrO$_3$ system. The distortion  parameter $\Delta$  of a coordination polyhedron BO$_N$ with an average bond length B-O $\langle$d$\rangle$, is defined as $\Delta$ = (1/N)$\sum_{n=1, N}$ \{(d$_n$-$\langle$d$\rangle$)/$\langle$d$\rangle$\}$^2$. The average tilt angle $\langle\phi\rangle$ of CrO$_6$ octahedral around the pseudocubic $<$111$>$  direction is obtained from the two angles; $\theta_1$ and $\theta_2$.}
\begin{ruledtabular}
\begin{tabular}{ccccccc}
x         & 0.0 & 0.025 & 0.05 & 0.075 & 0.10 & 0.15 \\
\hline
Cr-O(1):2 & 1.984(3)& 1.966(3) & 1.961(3) & 1.956(3)  & 1.957(2) & 1.956(3) \\
Cr-O(2):2 &1.980(3)&1.986(3)&1.986(2)&1.985(2)&1.986(3)&1.982(2)\\
Cr-O(2):2& 2.023(2)  & 1.996(3)&  1.989(3)&  1.987(3) &   1.985(3)    & 1.988(3)\\
 $\langle$Cr-O$\rangle$ &    1.995&   1.983&   1.979 &  1.976& 1.976& 1.975\\
$\Delta$(Cr-O)x10$^{-5}$ &  18.6 &   7.8& 8.0& 10.2&    9.1& 9.9\\
$\theta_1$=Cr-O(1)-Cr &148.48(2)&   146.77(2) &  146.60(3) &  148.44(2)&   149.94 & 148.43(3)\\
$\theta_2$=Cr-O(2)-Cr &146.21(2) & 146.10(2)& 146.51(3)& 146.64(2)&   147.99&  146.80(3)\\
$<\langle\phi\rangle$ &    20.0 &   20.5 &   20.4 &   19.8 &   19.0  &  19.8\\

\end{tabular}
\end{ruledtabular}
\end{table}

\subsection{Magnetic properties}

Figure 3 a) shows the temperature dependence of the inverse susceptibility, $\chi^{-1}$, from room temperature down to 2 K for x=0.0, 0.05, 0.1, and 0.2 of Ca content. At first sight, two features are clearly observed in the magnetization curves: firstly, there are not significant changes in the transition temperature which occur at about 140 K, and secondly, it is observed slight changes in the paramagnetic state (From N\'eel temperature up to 300 K). In order to further investigate the magnetic behavior of Ca-doped compounds, the Curie-Weiss law was fitted using the formula;

$\chi^{-1}= [C / (T -\theta)]^{-1}$,

\noindent
where $C$ is the molar Curie constant and $\theta$  is the Curie-Weiss temperature. The fit gives an $\mu_{eff}$ value for x=0 about 4.24 $\mu_B$ that is ~0.35 $\mu_B$ higher than that reported of 3.87 $\mu_B$ for Cr(d$^3$) with S=3/2 \cite{18}. We observed a slight increase in number of Bohr magnetons from 4.24 $\mu_B$ for x=0 to 4.30 for x=0.050. After that, the effective moments decrease to 4.22 $\mu_B$ and 4.06 for x=0.10 and 0.20, respectively. The experimental magnetic effective moment $\mu_{eff}$ and Curie-Weiss temperature as Ca is replaced in the Y sites are shown in Table III. The negative Curie-Weiss temperature ($\theta_{cw}$) is indicative of antiferromagnetic (AF) exchange interaction. In Fig. 3 b) we show the magnetization versus applied magnetic field recorded at 5 K for the pristine sample, and the doped with 20\% of Ca. The hysteresis loop is shown in both samples with linear decreasing starting from the maximum magnetic field (5 kOe) and down to H = 0 kOe. It is observed a slight increase of the remanent magnetization and coercive field when 20\% of Ca is replaced in the Y site. In addition, a more notable effect of Ca in the hysteresis curve is observed at high magnetic fields (40-50 kOe) where the magnetization increases from 0.072 for x = 0 to 0.087 for x = 0.020 $\mu_B$/f.u, respectively.

\begin{table}
\caption{\label{t3}Effective magnetic moment and Curie-Weiss temperature for Y$_{1-x}$Ca$_x$CrO$_3$.}
\begin{ruledtabular}
\begin{tabular}{ccc}
Ca          & $\mu_{eff}$ & $\theta_{W}$\\
Composition & ($\mu_B$) & (K) \\
\hline
0 & 4.24 & 401 \\
0.50 & 4.30 & 410 \\
0.10 & 4.22 & 390 \\
0.20 & 4.06 & 355 \\
\end{tabular}
\end{ruledtabular}
\end{table}

\begin{figure}
\includegraphics[scale=0.5]{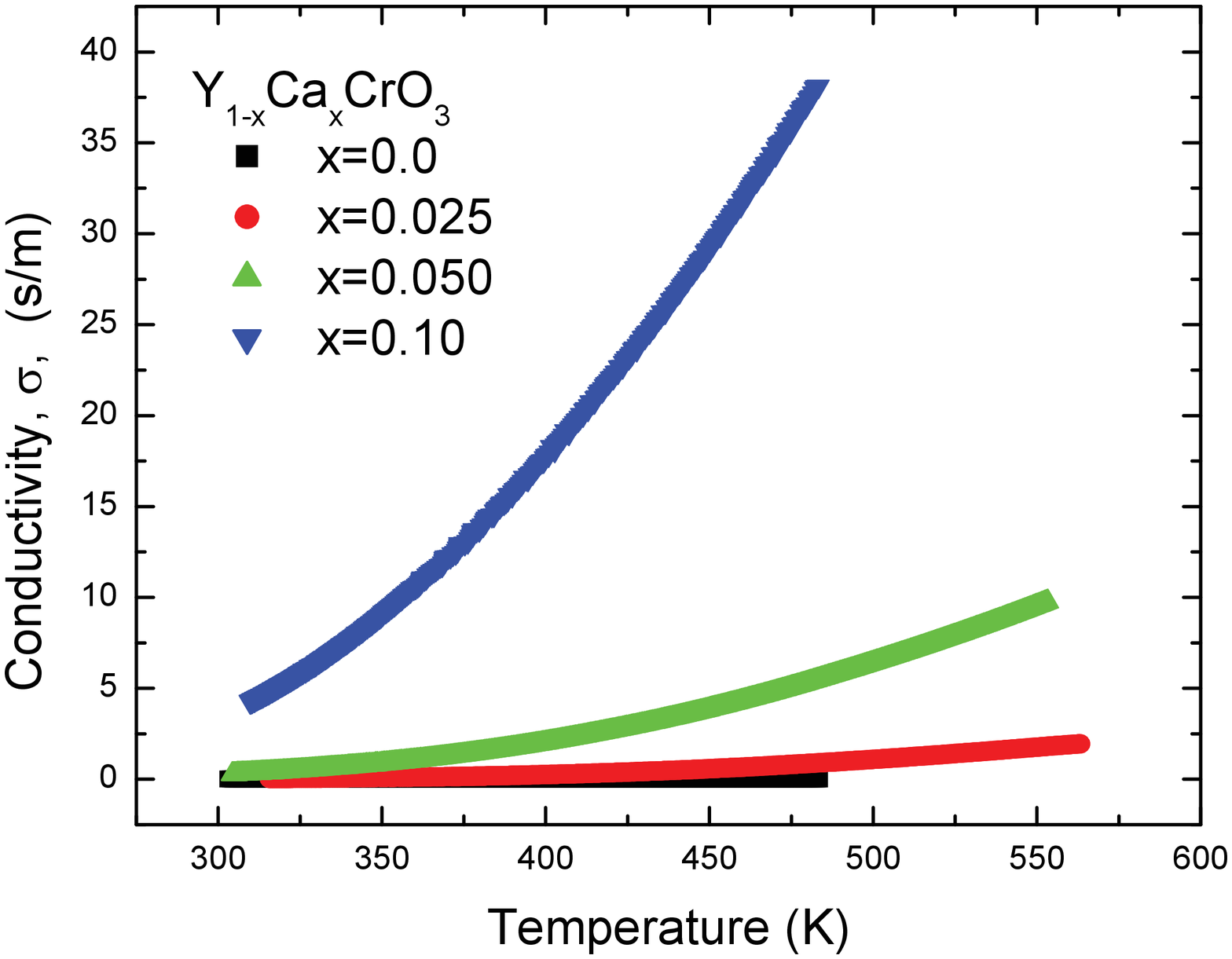}
\caption{\label{f4}Electrical conductivity ($\sigma$) as a function of temperature for Y$_{1-x}$Ca$_x$CrO$_3$ with $0 \leq x \leq 0.10$.}
\end{figure}

\subsection{Electric properties}

The effect of Ca on the transport properties was observed through measurements of the a.c. electrical conductivity ($\sigma$). This was determined from the capacitance and dielectric loss (tan$\delta$) vs temperature data at each value of frequency, $\omega_i$ using the formula:

$\sigma_{\omega i}(T)= \frac{l}{s}\omega_i Cp_{\omega_i}(T)Tan \delta_{\omega i}$.

In this formula ``$l$" is the thickness and ``$s$" the area of the electrode deposited on the sample. Conductivity as a function of temperature in the range $0 \leq x \leq 0.1$ of doped Ca is illustrated in the Fig. 4.  The curves clearly show that the doping of Ca increases the electrical conductivity in YCrO$_3$ matrix. In order to know the conduction mechanism and the associated activation energy values, E$_{act}$, in doped YCrO$_3$, we have fitted the electrical conductivity (at 10 kHz) to the Arrhenius law as:

$\sigma(T)=\sigma_0 e^{-(E_{act}/k_BT)}$

\noindent
here, $\sigma_0$ is the characteristic conductivity of the material in general dependent of the frequency, $k_B$ is the Boltzmann's constant, T is the absolute temperature and E$_{act}$ is the activation energy associated with the conduction mechanism in the region of temperature analyzed. This can be calculated from the slope of the ln$\sigma$ vs 1000/T as is shown in Fig. 5. The values of the conductivity at a selected temperature (373 K) and activation energy obtained from the fitting, are summarized in Table IV. At first instance, the collected data show that the conductivity at 373 K increases while the activation energy decreases as calcium is doping in the YCrO$_3$ compound. Another manner to make an interpretation of these results is that the system diminishes its activation energy, and increasing the transport process when calcium is introduced in the lattice; {\sl i.e.}, the charge carriers require less energy so that they can be thermally activated by hopping process as a transport mechanism.

\begin{figure}
\includegraphics[scale=0.5]{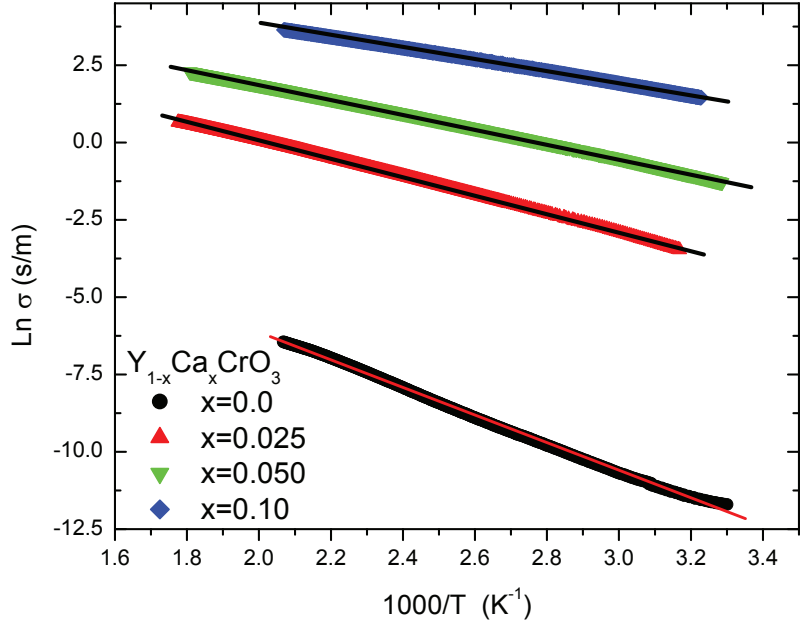}
\caption{\label{f5}Logarithm of conductivity ($\sigma$) as a function of the reciprocal temperature for Y$_{1-x}$Ca$_x$CrO$_3$ with $0\leq x \leq 0.10$. The fitting line is indicated.}
\end{figure}

\begin{table}
\caption{\label{t4}Electrical conductivity and activation energy for Y$_{1-x}$Ca$_x$CrO$_3$.}
\begin{ruledtabular}
\begin{tabular}{lll}
Composition & $\sigma_{100^\circ C}$  & E$_{act}$ \\
&(s/m)  &(eV)\\
\hline
0 & 9.589x10$^{-5}$ & 0.38 \\
0.025 & 0.962 & 0.26 \\
0.50 & 1.307 & 0.21 \\
0.10 & 12.55 & 0.17 \\
\end{tabular}
\end{ruledtabular}
\end{table}

\section{Discussion}

\subsection{Structure and magnetization analysis}

It seen in Fig. 2 that the unit cell volume decreases as Ca content increases. This behavior is contrary to the expected result, considering the difference of ionic radii \cite{19} with eight coordination number between Ca (1.12 \AA) and Y (1.019 \AA) suggesting that the c-parameter and cell volume should increase as Ca is replaced at the Y site. The decrease of cell volume may be attributed to the charge increase from Cr(III) to Cr(IV) in the octahedral environment. Thus, the cell volume decreasing may be explained by the change of Cr(III) with ionic radii of 0.615 \AA\ to Cr(IV) with ionic radii of 0.55 \AA\ \cite{19}. This fact is supported by the continuous decreasing of the $\langle$Cr-O$\rangle$ interatomic distance from 1.995 to 1.975 \AA\ for x=0.15 of Ca (see Table II). Recently, Arévalo-López et al. \cite{20} have correlated the oxidation state, average $\langle$Cr-O$\rangle$ distance and energy difference between the CrL$_3$ and O K edges using electron energy loss spectroscopy (EELS) for several Cr(III) and Cr(IV) compounds. Here, the $\langle$Cr-O$\rangle$ distance for x=0 agrees well with the valence of Cr(III) in Cr$_2$O$_3$ and according to these collected data, the average $\langle$Cr-O$\rangle$ distance should achieve 1.908 \AA for Cr(IV) in the end CaCrO$_3$ composition \cite{21}. These subtle structural changes should take effect on the magnetic properties.

Taking into consideration that neither the yttrium nor calcium contribute to the magnetic behaviour, the slight deviation of the effective moments for x = 0 could be due to the magnetic fluctuations persisting in the range of fitting temperature. The slight increase in number of Bohr magnetons and Curie-Weiss temperature (see table II) after doping with Ca could be the result of a crystal field perturbation as a consequence of the charge rearrangement on the Cr-environment, resulting by the charge difference between replacement  of Ca(II) by Y(III). This fact is supported by structural consideration as it is observed in the cell volume increase and the octahedral distortion ($\Delta$) drops, from 18.8 to 8.0 (see Table II) when is introduced 2.5 and 5.0 \% of Ca into the YCrO$_3$ matrix. Now, if the effect of doping with Ca$^{+2}$ into the YCrO$_3$ matrix is to produce the corresponding number of Cr$^{+4}$ ions {Cr(d$^2$) with S=1, $\mu_{eff}$=2.83  $\mu_B$, as is seen experimentally by structural consideration, then we should anticipate a decrease in the number of Bohr magnetons. Such effect is experimentally observed after x = 0.050 of Ca content (Table II). However, it is important to point out that magnetic measurements at higher temperatures would be required to cancel magnetic fluctuations and to obtain more reliable information of the effective moments for the ground state of Cr(d$^3$) with S =3/2 \cite{22}. On the other hand, the M(H) curves confirm the weak ferromagnetism by canted antiferromagnetism, which is due to antisymmetric Cr-Cr spins interactions, in the AF G-type magnetic structure \cite{23,24,25}. It has been reported that the remanent magnetization and coercive field decreases to the end composition of CaCrO$_3$ \cite{21}. Oxidation states changes from Cr(III) to Cr(IV) in Y$_{1-x}$Ca$_x$CrO$_3$ compounds must be reflected in the electrical transport properties since the Cr(IV) in many oxides tends to behave as a semi-metal, as in particular CaCrO$_3$ which is an itinerant-electron antiferromagnetic insulator \cite{25,26,27}.

\begin{figure}
\includegraphics[scale=0.5]{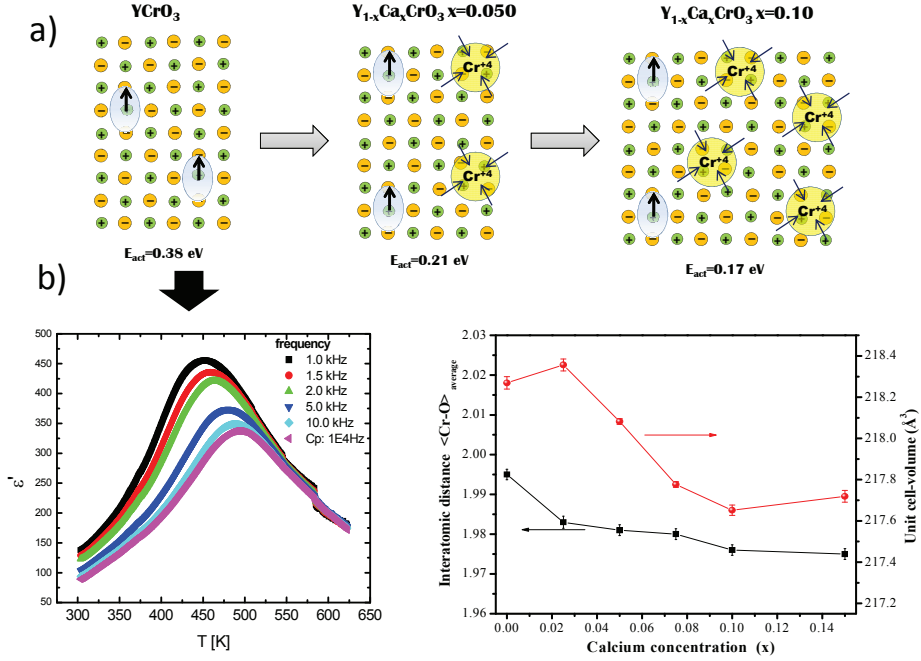}
\caption{\label{f6}a) Mechanism of small-polaron formation. Deformation field as result of ``local non-centrosymetry" is the electron-phonon interaction responsible of the small-polarons formation, see as ``lobe with arrows", in undoped sample, while the change from Cr(III) to Cr(IV)  is the responsible of small-polarons formation, see as ``circle with arrows", in doped samples. b) Curve of permittivity vs temperature where is shown the ferroelectric-paraelectric transition as a consequence of the local lattice deformation. The decrease of the cell volume and interatomic $\langle$Cr-O$\rangle$ average distance is a clear evidence of the Cr(IV) formation, responsible of small-polarons, as Ca is doped in the YCrO$_3$ matrix.}
\end{figure}

\subsection{Transport analysis and mechanism of small-polarons formation}

Several conduction mechanisms associated with activation energy values have been reported for empty d shell or ferroelectric materials [Raymond {\sl et al}. \cite{28} and reference therein]. Also, several ortochromites exhibit thermally activated conductivity with smaller activation energies \cite{29}. In many of these, the positive Seebeck coefficient indicates p-type conductivity, where the carrier transport has been associated to small-polarons. In the case of Ca doped in YCrO$_3$, the E$_{act}$ values calculated for each Ca concentration (see Table IV) correspond to the small-polarons, in which, their associated E$_{act}$ is reported from 0.21 to 0.8 eV for perovskite ceramic samples \cite{28}. For instance, the E$_{act}$ values for x=0 is about 0.1 eV, higher than the value of 0.24 eV reported by Rao et al. in single crystals \cite{27} and the values of E$_{act}$ for doped samples reported here are in good agreement with these results reported by Weber et al. in polycrystalline samples \cite{30}. Microstructural features resulting from the synthesis method may be a main factor that influences the deviation of the E$_{act}$ for undoped sample to those reported in the literature \cite{30,31}. Nevertheless, the decrease of E$_{act}$ from 0.38 to 0.17 eV as Ca doping is associated to the formation of small-polarons as charge carriers. It is known that cations with low oxidation state doping in Y or La-chromites trigger small-polarons as charge carrier \cite{30,32}. In Fig 6, an empirical mechanism, using experimental evidence is proposed for undoped and doped Y$_{1-x}$Ca$_x$CrO$_3$. The origin of small-polaron has been uncertain in pure YCrO$_3$ since the conditions for small-polarons formation are a narrow conduction band and a large electron-phonon interaction \cite{33,34}. The lattice distortion field necessary for the small-polaron formation is absent in pristine samples. Earlier study \cite{35} in undoped samples assumed that the inherent defects, oxygen vacancies, lead to the formation of Cr(IV) identical to those formed by Ca-doped. These facts are not convincing to explain the origin of the small-polarons in undoped samples. The charge compensation through oxygen vacancies as well as the Cr(IV) creations in the lattice should contribute to the conductivity data as a change of slope in the ln$\sigma$ vs 1/T plot since both must be activated at different temperature. Conduction mechanisms through oxygen vacancies have been reported \cite{28,36} with E$_{act}$ from 0.64 to 0.89 eV. This conduction mechanism with that E$_{act}$ is not observed in Fig. 5. Local non-centrosymmetric structure recently reported by Ramesha et al. \cite{11} is the key to understand the formation of small-polarons as a charge carriers for undoped samples. The Cr displacements of about 0.01 \AA\ along the z direction are clear evidence of a local polarization field. In Fig. 6 a), we show a mechanism for trigger charge transport as a small-polarons in undoped and doped YCrO$_3$ compound. For undoped sample, the local polarization field is responsible for the weak ferroelectricity observed at 450 K in the permittivity vs. temperature curve in Fig. 6 b). In addition, we believe that this condition is necessary for small-polarons formation, which allow charge carriers locally around the Cr displacements (lobe with arrow) as is seen in Fig. 6 a) for YCrO$_3$. The carrier, together with the distortion field, then, move by thermal activated hopping. As Ca is doping in the lattice, the charge compensation as a consequence of different substituting valence (Ca by Y) occurs through local change from Cr(III) to Cr(IV). This fact increases the local polarization field. Thus, the decrease of the E$_{act}$ and consequently the increase of the electrical conductivity are result of the formation of more local polarization field as is seen for Y$_{1-x}$Ca$_x$CrO$_3$ with x = 0, 0.05 and 0.1 in Fig. 6 a). It is important to note that the change from Cr(III) to Cr(IV) ions in the lattice is experimentally supported by structural and magnetic measurements. Here, the continuous decrease of the $\langle$Cr-O$\rangle_{AVG}$ interatomic distance and cell volume (Fig. 6 b) as well as the decrease of the effective magnetic moment is clear evidence that the Ca doping alter the charge of the YCrO$_3$ through local changes in the octahedral environment in the lattice (local non-centrosymetry plus Cr(IV) formation). This, in turns, gives rise to carrier injection as small-polarons. These quasi-particles move from site to site by thermally activated hopping contributing to the conduction between room temperature and $\sim$473 K. It is important to remark that the conductive process induced by charge fluctuation in the transition metal is detrimental to the ferroelectric behavior, which in many cases; this intrinsic characteristic (semiconducting behavior) give rise to leakage currents, which interfere with polarization switching and pooling process in the biferroic materials.

\section{Conclusions}

We studied the behavior of structural, magnetic and electrical conductivity in YCrO$_3$ doped with Ca. The continuous decreasing of volume cell as well as the average $\langle$Cr-O$\rangle$ bond length is an indicative of the increasing of the average formal oxidation state of the Cr-environment. We did not find significant changes in the antiferromagnetic transition temperature, about 140 K. Contrary to this behaviour significant changes were observed in the electrical transport properties. The activation energy extracted by Arrhenius law decreases as a consequence of the formation of small-polarons as Ca is doped. Local non-centrosymetric distortion field and charge compensation through the change from Cr(III) to Cr(IV) giving raise to small-polarons formation, mainly localized in the Cr(III)-Cr(IV) sites. The decrease of the average $\langle$Cr-O$\rangle$ interatomic distance and the effective moments in the paramagnetic state, as well as the increase of the electric conductivity with Ca support these results. These findings in bulk multiferroics materials, with partially filled $d$-orbitals, are detrimental to the optimum performance of the magnetoelectric properties.

\begin{acknowledgments}
A. D. thanks to DGAPA-UNAM through project IN112909 and also the Scientific Coordination-UNAM in the program ``Intercambio Acad\'emico". RE thanks DGAPA-UNAM project IN100711, CONACyT financial support, to BISNANO, and to the Instituto de Ciencias del Departamento del D. F., Cd. de M\'exico. FM thanks the partial support of DGAPA-UNAM project IN111511. We also thank to I. Gradilla (CNyN), E. Aparicio (CNyN), J. Palomares (CNyN), P. Casillas (CNyN) and E. Flores (CNyN) for their technical help.
\end{acknowledgments}

\end{document}